\documentclass{article}
\usepackage{ijcai20}

\usepackage{times}

\usepackage{soul}
\usepackage{url}
\usepackage[hidelinks]{hyperref}
\usepackage[utf8]{inputenc}
\usepackage[small]{caption}
\usepackage{graphicx}
\usepackage{subcaption}
\usepackage{caption}
\usepackage{siunitx}
\usepackage{multirow}
\usepackage[T1]{fontenc}
\usepackage{amsmath}
\usepackage{booktabs}
\usepackage{color}

\usepackage{pdflscape}

\title{Deep Learning in Medical Ultrasound Image Segmentation: a Review}

  \author{
 Ziyang Wang$^1$\footnote{Contact Author}
 \\
 \affiliations
 $^1$University of Oxford\\
 \emails
 {ziyang.wang}@cs.ox.ac.uk
  }

\begin{document}

\maketitle
\begin{abstract}
Applying machine learning technologies, especially deep learning, into medical image segmentation is being widely studied because of its state-of-the-art performance and results. It can be a key step to provide a reliable basis for clinical diagnosis, such as the 3D reconstruction of human tissues, image-guided interventions, image analyzing and visualization. In this review article, deep-learning-based methods for ultrasound image segmentation are categorized into six main groups according to their architectures and training methods at first. Secondly, for each group, several current representative algorithms are selected, introduced, analyzed and summarized in detail. In addition, common evaluation methods and datasets for ultrasound image segmentation are summarized. Furthermore, the performance of the current methods and their evaluation results are reviewed. In the end, the challenges and potential research directions for medical ultrasound image segmentation are discussed.
\end{abstract}

\section{Introduction}
\begin{figure}
     \centering
     \begin{subfigure}[b]{0.21\textwidth}
         \centering
         \includegraphics[width=\textwidth]{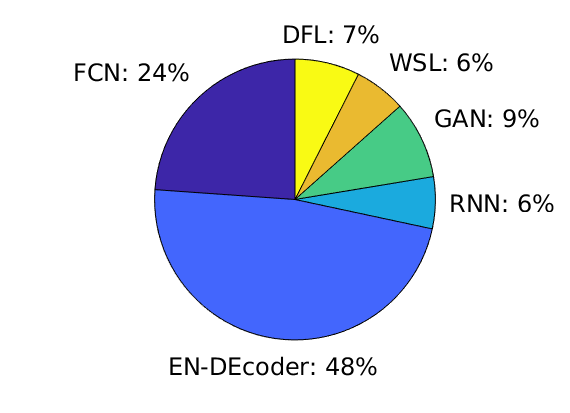}
         \caption{The percentage of number of papers for each group}
         \label{fig:piechart}
     \end{subfigure}
     \hfill
     \begin{subfigure}[b]{0.21\textwidth}
         \centering
         \includegraphics[width=\textwidth]{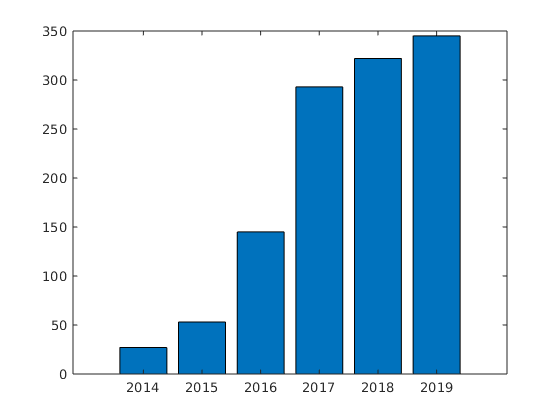}
         \caption{The number of published papers for US segmentation}
         \label{fig:barchart}
     \end{subfigure}
        \caption{Overview of published papers}
        \label{fig:overview}
\end{figure}

Medical ultrasound imaging provides the inside structure of the human body with high-frequency sound waves which is safe, painless, noninvasive, non-ionized and real-time. Ultrasound imaging as compared to other imaging tools, such as Computed Tomography (CT) and Magnetic Resonance Imaging (MRI), is cheaper, portable and more prevalent. It helps to diagnose the causes of pain, swelling, and infection in internal organs, for evaluation and treatment of medical conditions. Ultrasound imaging has turned into a general checkup method for prenatal diagnosis. However, several drawbacks such as the necessity of a skilled operator, the difficulty of distinguishing imaging structures between tissue and gas, and the limitation of the field of view bring more challenges on image processing algorithms study. Many approaches have been proposed to solve these problems, and one of the most important approaches is deep learning. The development of deep-learning-based medical ultrasound image segmentation technology plays an essential and fundamental role in the analysis of biomedical images and significantly contributes to classification, recognition, visualization, 3D reconstruction and image-guided intervention, which can provide reliable guidance for doctors in clinical diagnosis.
% \cite{garcia2017review}\cite{guo2018review}\cite{huang2017review}\cite{liu2019deep}\cite{noble2006ultrasound}\cite{tajbakhsh2019embracing}\cite{xian2018automatic}\cite{zhou2019review}\cite{taghanaki2019deep}\cite{seo2019machine}\cite{chen2019deep}\cite{van2019deep}\cite{DBLP:journals/jdi/HesamianJHK19}\cite{liu2019deep}\cite{hacihaliloglu2017ultrasound}
Compared with previous papers on image segmentation and deep learning, this is the first review on applying deep learning approaches to medical ultrasound image segmentation. The following contributions are made for deep learning and ultrasound image segmentation community:
\\$\bullet$ To the best of our knowledge, this review paper is the only paper provides comprehensive introduction and memorization on medical ultrasound image segmentation with deep learning methods.   
\\$\bullet$ This review paper groups the related deep learning literature into six sections based on different architectures or training method of deep learning models including: Fully Convolutional Neural Networks (FCN), Encoder-Decoder Neural Networks (EN-DEcoder), Recurrent Neural Networks (RNN), Generative Adversarial Networks (GAN), Weakly Supervised Learning (WSL) and Deep Reinforcement Learning (DFL) methods. The overview of the number of published papers in recent years is shown in Figure \ref{fig:overview}.
\\$\bullet$ The evaluation methods, common datasets and experimental performance results of current deep learning approaches are systematically organized and reviewed.

The following Section \ref{section:FCN}, Section \ref{section:EDN}, Section \ref{section:RNN}, Section \ref{section:GAN}, Section \ref{section:WSL}, and Section \ref{section:RFL} review each categorized group of deep learning models, respectively. Section \ref{section:evaluation} summarizes common evaluation methods, datasets and the comparison of results. Section \ref{section:DISCUSSION} summarizes and discusses the key issues and potential research directions in the field of deep learning in ultrasound image segmentation.  

\section{Fully Convolutional Neural Networks for Segmentation}\label{section:FCN}
Fully Convolutional Networks (FCN) is firstly introduced by \cite{long2015fully}, which is one of the most commonly used CNN-based neural networks for image semantic segmentation. It is a pixels-to-pixels supervised learning approach which is compatible to any size of images. The FCN architecture shown in Figure \ref{fig:FCN} is mainly developed by a classification network where fully connected layers are replaced with fully convolutional networks. A skip step (skip layer and bilinear interpolation) is proposed to extend the application of classification network to dense prediction, so that accurate pixel-level segmentation is achieved on the entire images.\

Meanwhile, several state-of-the-art FCN based methods have been proposed in medical ultrasound image semantic segmentation. \cite{zhang2016coarse} proposed a lymph node segmentation framework in 2016. The framework consists of two FCN-based modules which the first one is to collect raw images to produce potential objects' segmentation, and the other one generates final lymph nodes segmentation with the intermediate results and the raw images. Multi-stage incremental learning and multi FCN modules concept are introduced and investigated. The CFS-FCN model is shown in Figure \ref{fig:CFS-FCN}\

\cite{wu2017cascaded} came up a cascaded fully convolutional networks for ultrasound image segmentation in 2017. To avoid the boundary incompleteness and ambiguity, an Auto-Context scheme modified by join operator is implanted into FCN model. \cite{mishra2018ultrasound} proposed a FCN model with attentional deep supervision named DSN-OB in 2018. Coarse resolution layers are trained to discriminate the object region from background, and fine resolution layers are trained for object boundary definitions. A specific training scheme and fusion layer are developed to avoid the broken boundaries which is a common ultrasound image segmentation challenge. The framework is finally approved a promising performance for blood region and lesion segmentation, respectively. 

In addition, 3D Ultrasound imaging can provide richer spatial information of the tissues than the traditional 2D X-ray imaging. To better exploit the 3D spatial information while making full use of the 2D pretrained model, Yang et al. proposed a catheter detection method named Direction-Fused FCN (DF-FCN), which exploits 3D information through re-organized cross-sections to segment the catheter\cite{yang2019improving}. In this work, the 3D ultrasound volume is divided into several 2D images firstly, then each image is processed by the FCN to acquire the probability prediction. Moreover, as shown in Figure \ref{fig:DF-FCN}, the results predicted by FCN are stacked together on the original position to construct the feature maps in three different directions, and these three feature maps are direction-fused to enhance the inter-slice context information. At last, the final segmentation results are obtained by applying a 3D convolution and a softmax layer.

% \cite{brown2019deep}\cite{dhindsa2018grading}\cite{cunningham2019ultrasound}\cite{ma2017cascade}\cite{villa2018fcn}\cite{zhang2018fully}\cite{sundaresan2017automated}\cite{zhang2016coarse}\cite{ma2017ultrasound}\cite{singhal2017automated}\cite{yu2016segmentation}\cite{cheng2017transfer}\cite{menchon2015fully}

\section{Encoder-Decoder Neural Networks for Segmentation}\label{section:EDN}
To recover the pixels' location information lost in pooling operation, encoder-decoder architecture based networks were proposed introducing opposite operations including convolution and deconvolution (or transpose convolution), pooling and unpooling. The encoder approach is to collect pixel location features, and then decoder approach is to restore the spatial dimension and pixel location features. Therefore, the information of feature and pixel location can be fully retained and analyzed. 

In 2015, a completely symmetric architecture, DeconvNet was proposed by \cite{noh2015learning} with deconvolution and unpooling layers. It was applied to ultrasound segmentation of cervical muscle in \cite{cunningham2019ultrasound}, comparing performance of integrating with Exponential Linear Unit (ELU) and Maxout to solve `dying ReLU’ problem. 

\begin{figure*}
\centering
\begin{subfigure}[b]{0.22\textwidth}
\centering
    \includegraphics[width=\textwidth]{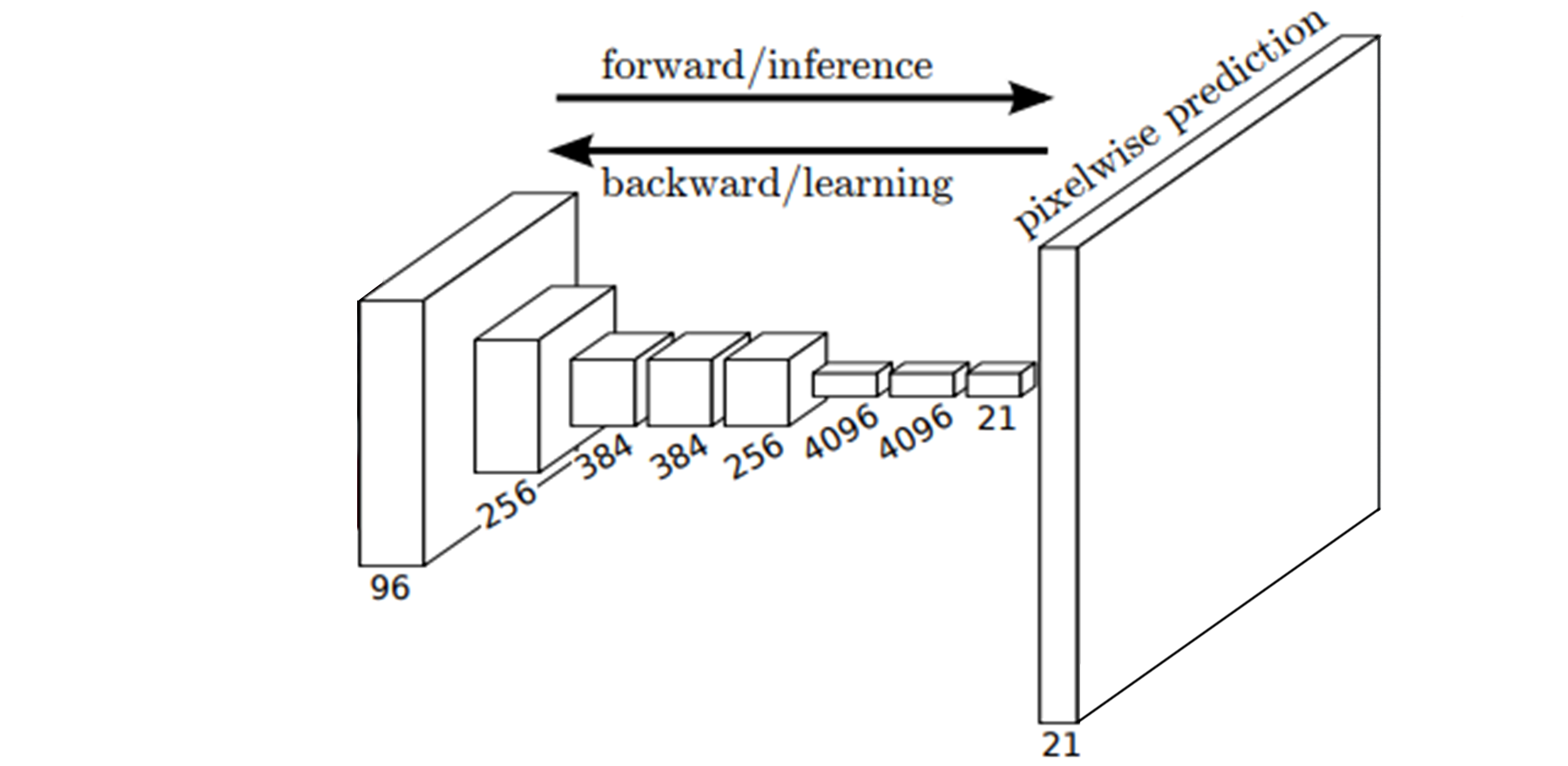}
    \caption{FCN}
    \label{fig:FCN}
\end{subfigure}
% \hfill
\begin{subfigure}[b]{0.22\textwidth}
\centering
    \includegraphics[width=\textwidth]{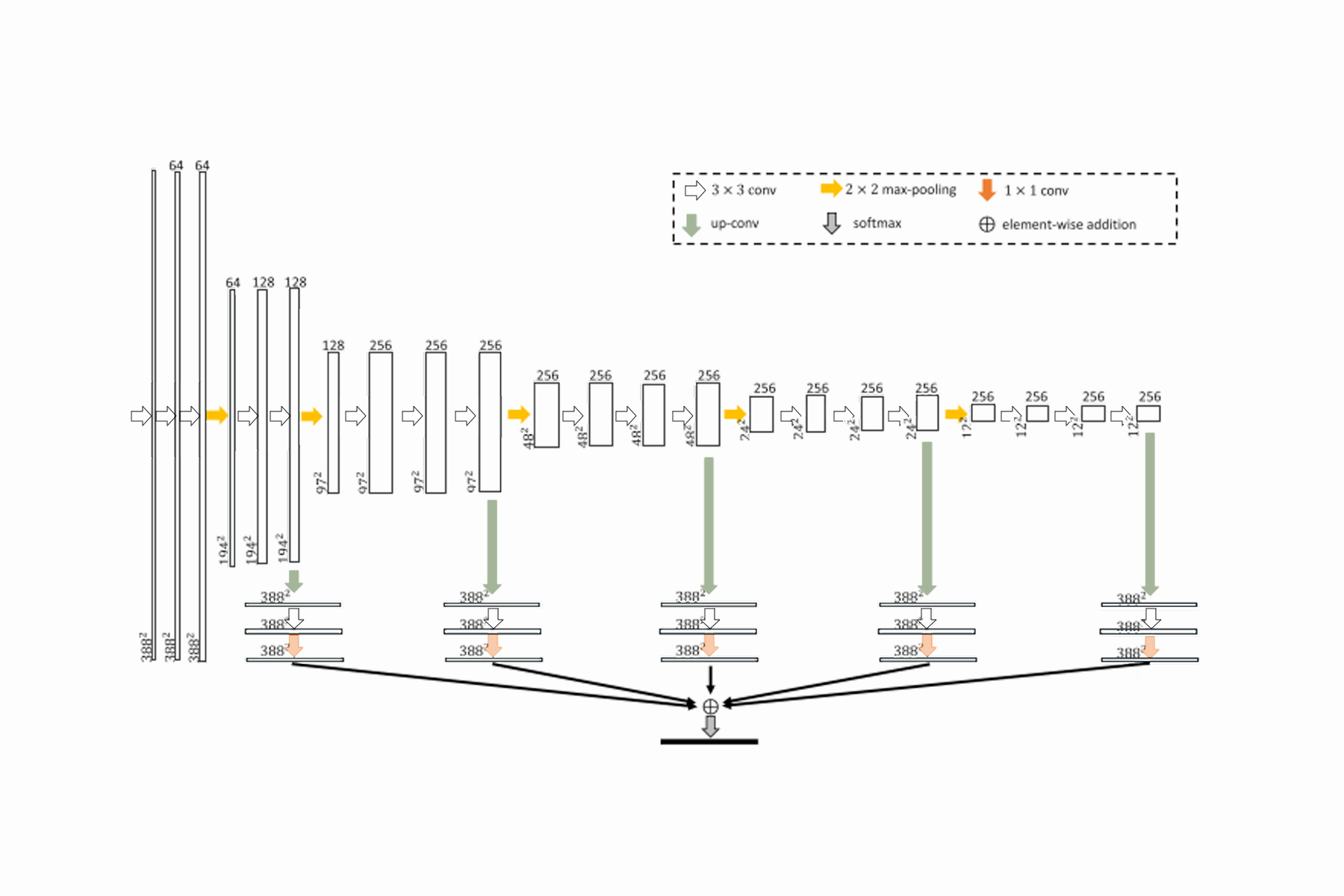}
    \caption{CFS-FCN}
    \label{fig:CFS-FCN}
\end{subfigure}
% \hfill
% \hfill
\begin{subfigure}[b]{0.32\textwidth}
    \centering
    \includegraphics[width=\textwidth]{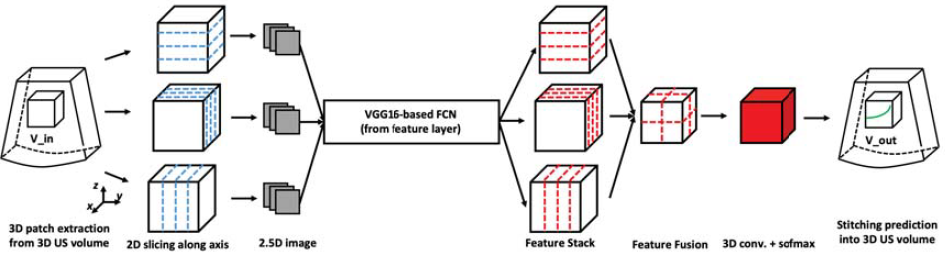}
    \caption{DF-FCN}
    \label{fig:DF-FCN}
\end{subfigure}
\begin{subfigure}[b]{0.22\textwidth}
    \centering
    \includegraphics[width=\textwidth]{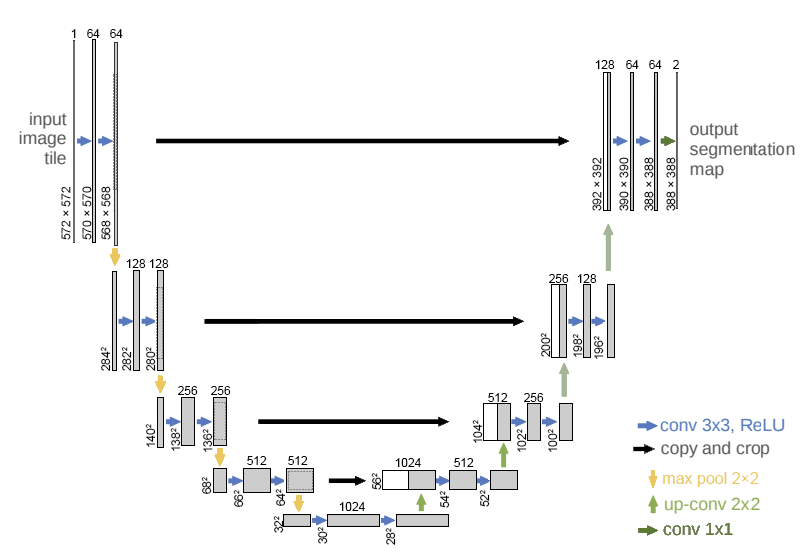}
    \caption{U-Net}
    \label{fig:UNET}
\end{subfigure}
\begin{subfigure}[b]{0.22\textwidth}
    \centering
    \includegraphics[width=\textwidth]{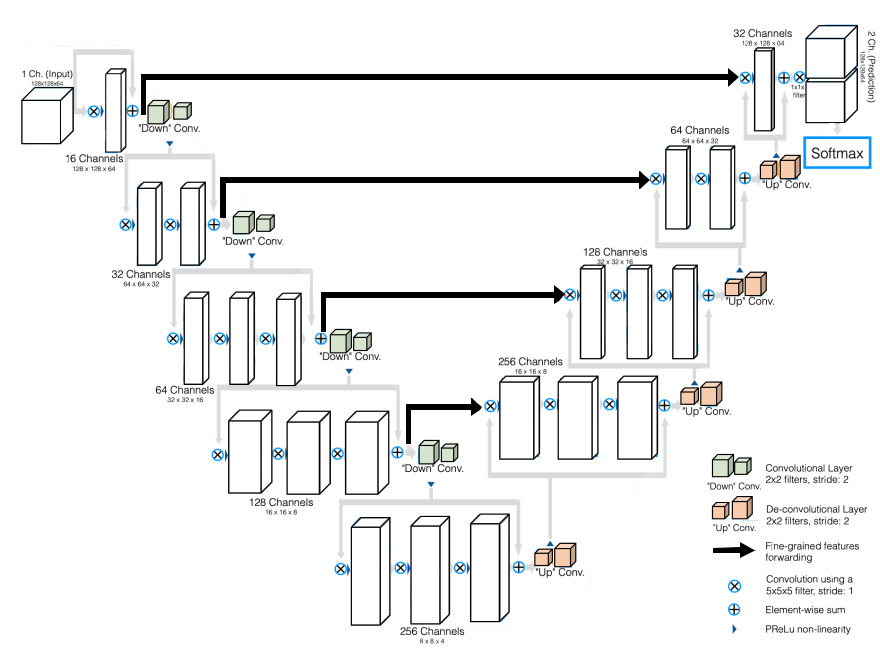}
    \caption{V-Net}
    \label{fig:V-Net}
\end{subfigure}
% \hfill
% \hfill
% \hfill
% \hfill
\begin{subfigure}[b]{0.25\textwidth}
    \centering
    \includegraphics[width=\textwidth]{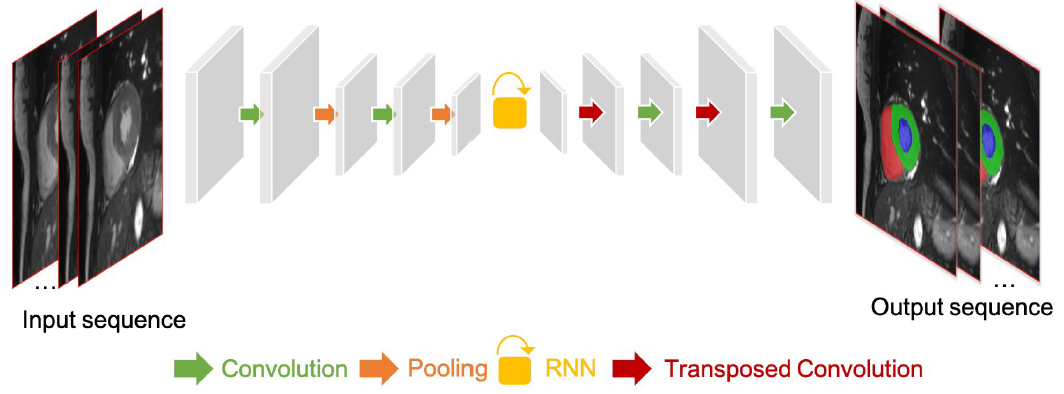}
    \caption{RNN}
    \label{fig:ReSeg}
\end{subfigure}
\begin{subfigure}[b]{0.22\textwidth}
    \centering
    \includegraphics[width=\textwidth]{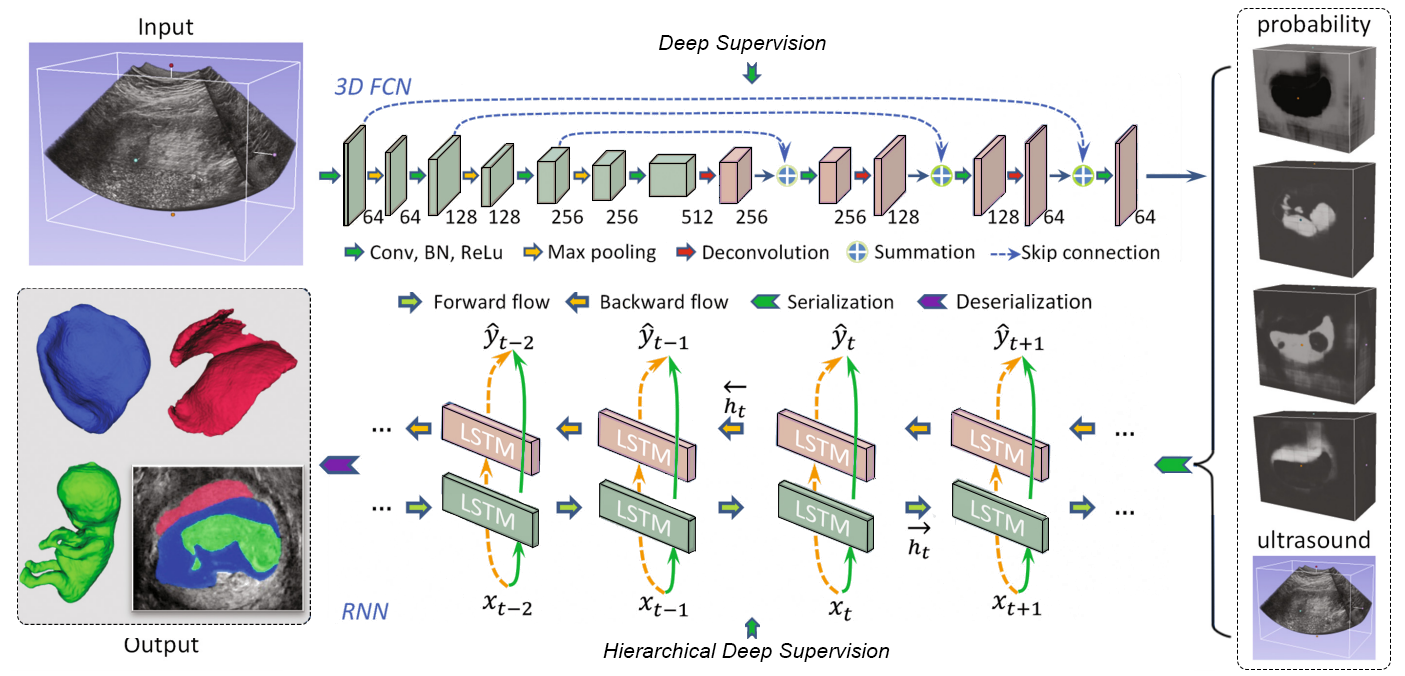}
    \caption{3D FCN and RNN}
    \label{fig:volume us}
\end{subfigure}
\begin{subfigure}[b]{0.22\textwidth}
    \centering
    \includegraphics[width=\textwidth]{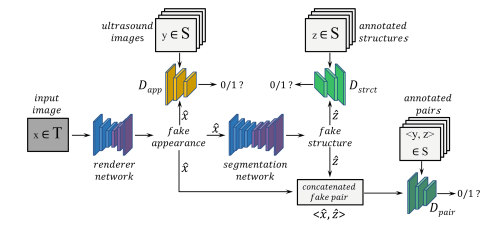}
    \caption{GAN}
    \label{fig:yang2018}
\end{subfigure}
% \hfill
\caption{The architecture for common deep learning neural networks}
\label{fig:fig}
\end{figure*}

In the same year, \cite{ronneberger2015u} proposed U-net, as shown in Figure~\ref{fig:UNET}, with increased channel number in decoder for propagation of context feature in higher resolution layers, and cropped feature maps concatenated from encoder to decoder for localizing each pixel. Introducing short connection into encoder part of U-net to avoid overfitting and dynamically fine tune in specific test task, Zhou et al. proposed a dynamic convolution neural network for media-adventitia and lumen-intima segmentation in ultrasound \cite{zhou2019deep}. \cite{zhuang2019nipple} proposed Grouped-Resaunet (GRA U-net) for slice-by-slice nipple segmentation and localization on breast ultrasound. GRA U-net is an architecture based on U-net, employing residual block for solving vanishing gradient, group convolution for computational efficiency and attention gates for focusing on relevant areas of input. Kim et al. proposed a network with the multi-scale input and hybrid multi-label loss function for segmentation of coronary arteries in intravascular ultrasound image \cite{kim2018fully}. Neural Architecture Search (NAS) was applied to semantic segmentation network in \cite{weng2019unet} based on the structure of U-net, which was applied on a nerve ultrasound dataset.

After the success in 2D medical image segmentation achieved by U-net, \cite{cciccek20163d} first extended U-net to 3D architecture for volumetric segmentation with 3D convolution. In the same year, V-net shown in Figure \ref{fig:V-Net} was proposed by Milletari et al. for volumetric segmentation, integrated with 3D convolution and residual blocks \cite{milletari2016v}. Lei et al. introduced a multi-directional deeply supervised V-net, and applied it to prostate segmentation in ultrasound \cite{lei2019ultrasound}. The network predicts different resolution segmentation from each stage of decoder part, uses multi-directional contour refinement processing for fusion of segmentation and applies a stage-wise hybrid loss function to reduce convergence time. 
% \begin{figure}[h]
%     \centering
%     \includegraphics[width=8cm]{4.png}
%     \caption{XXX Neural Networks}\cite{zhuang2019nipple}
%     \label{fig:my_label}
% \end{figure}

% \begin{figure}[h]
%     \centering
%     \includegraphics[width=8cm]{5.png}
%     \caption{XXX Neural Networks}\cite{degel2018domain}
%     \label{fig:my_label}
% \end{figure}

% \begin{figure}[h]
%     \centering
%     \includegraphics[width=8cm]{6.png}
%     \caption{XXX Neural Networks}\cite{yang2018ivus}
%     \label{fig:my_label}
% \end{figure}

% \begin{figure}[h]
%     \centering
%     \includegraphics[width=8cm]{7.png}
%     \caption{XXX Neural Networks}\cite{nandamuri2019sumnet}
%     \label{fig:my_label}
% \end{figure}

% \cite{sobhaninia2019fetal}

% \cite{azzopardi2017automatic}\cite{behboodi2019ultrasound}\cite{zhou2019deep}\cite{bonmati2018automatic}\cite{zhou2019u}\cite{looney2018fully}\cite{almajalid2018development}\cite{weng2019unet}\cite{kim2018fully}\cite{thomson2019hepatic}\cite{8580214}\cite{yang2017towards}\cite{lei2019ultrasound}\cite{loram2019automated}\cite{ghavami2018automatic}\cite{kiraly2017deep}\cite{ravishankar2017learning}\cite{li2017automatic}\cite{jaumard2016tongue}\cite{li2017automatic}\cite{baka2017ultrasound}\cite{ravishankar2017learning}\cite{badrinarayanan2017segnet}\cite{noh2015learning}\cite{chen2018encoder}
%%%%%%%%%%%%%%%%%%%%%%%%%%%%%%%%%%%%%%%%%%%%%%%%%%%%%%%%%%%%%%%%%%%%%%
%%%%%%%%%%%%%%%%%%%%%%%%%%%%%%%%%%%%%%%%%%%%%%%%%%%%%%%%%%%%%%%%%%%%%%%

\section{Recurrent Neural Networks for Segmentation}\label{section:RNN}
To extract image historical context information and global features, Recurrent Neural Networks (RNN) is designed to handle sequential data, where the past learned knowledge can help make present decisions through a recurrent path. Long Short-Term Memory (LSTM) and Gated Recurrent Unit (GRU) are two most widely used architectures in the family of RNN, which can enable the network to model a long-term memory. As shown in Figure \ref{fig:ReSeg}, given a sequence of data, such as continuous 2D scan slices, a RNN model takes the first image as input, extracts the features to make prediction on each pixel and memorizes these information. Then, these information are propagated to help make decisions for the next image. In this way, the context information from adjacent slices can be fully utilized to improve the segmentation performance. 

\cite{chen2015automatic} proposed a novel framework for automatic fetal ultrasound standard plane detection, which is a hybrid model integrating CNN and RNN. A ROI detector is firstly trained via the joint learning of convolutional neural networks (J-CNN) to mitigate the challenge of limited annotated data. Then, the LSTM model is used to explore the spatio-temporal features based on the ROIs in consecutive frames. Finally, the standard planes are acquired if the prediction score is larger than a threshold. 

\cite{yang2017towards} applied a general 3D FCN to achieve dense voxel-wise semantic segmentation in ultrasound volumes, including fetus, gestational sac and placenta. He introduced a RNN trained with hierarchical deep supervision (HiDS) through leveraging the 3D context information between sequential information flows from different directions to refine the local segmentation results. The overall architecture of the network is shown in Figure \ref{fig:volume us}. In addition, \cite{yang2017fine} proposed a novel framework for automatic prostate segmentation in ultrasound images. He introduced a Boundary Completion Recurrent Neural Networks (BCRNN) to learn the shape prior with the biologically plausible and exploit these information to infer the incompleteness along prostate boundary which can help improve the segmentation performance. Then, multi-view predictions are fused to obtain a comprehensive prediction. Finally, a multiscale Auto-Context scheme is applied to refine the details of the final predictions. 

\section{Generative Adversarial Networks for Segmentation}\label{section:GAN}
Generative Adversarial Network (GAN) and their extensions have attracted a great deal of attention recently relying on the ability of tackling well known and challenging medical image analysis problems such segmentation, reconstruction, classification or data simulation \cite{kazeminia2018gans}. The general idea of GAN is the combination of two neural networks with opposing goals and use adversarial training to realize joint optimization. The first network is a generator network and jointly optimized with a discriminator as the second network. The goal of the generator is to generate outputs which can not be distinguished from a dataset of real examples by the discriminator network \cite{wolterink2018generative}. 

\cite{luc2016semantic} presented the first application of adversarial training to semantic segmentation, which can detect and correct higher-order inconsistencies between the segmentation results and ground truth segmentation maps. Their approach enforced long-range spatial label contiguity without increasing the model complexity. In the image segmentation, dense and pixel-level labeling are needed. It is challenging to produce sufficient and stable gradient feedback to the networks from the single scalar classic GAN's discriminator output. To overcome this limitation, a novel end-to-end adversarial neural network with a multi-scale L1 loss function was proposed in \cite{xue2018segan} for medical image segmentation task. In this framework, the segmentor and critic networks  were trained to learn both global and local features, which can capture long-range and short-range spatial relationships among pixels. 
For medical imaging, it is difficult to capture 3D semantics in an effective way while keep computation efficient at the same time. To address this, \cite{khosravan2019pan} showed a novel projective adversarial network `PAN', which represented high-level 3D information through 2D projections. In this framework, He also introduced an attention module to realize a selective integration of global information from segmentor to adversarial network directly. For automated ultrasound image segmentation, when come across appearance discrepancy, the deep models tend to perform poorly even on congeneric corpus where objects have similar structure but slightly different appearance. \cite{yang2018generalizing} tried to solve this general problem by using a novel online adversarial appearance conversion method shown in Figure \ref{fig:yang2018}. They put forward with a self-play training strategy to pre-train all the adversarial modules for acquiring the structure and appearance distributions of source corpus. In addition, the composite constraints for appearance and structure in the framework also helped to remove appearance discrepancy iteratively in a weakly-supervised model. 

% \cite{yasarla2019learning}\cite{son2017retinal}

\section{Weakly Supervised Learning for Segmentation}\label{section:WSL}

The manually pixel-level labeled image dataset providing a lot of detailed information can significantly improve supervised training efficiency and segmentation accuracy. Compared with the supervised training for CNN-based network, some researchers study on weakly supervised learning, because weakly labeled data requires low cost in labeling. 

\cite{kim2016deconvolutional} proposed a weakly supervised framework with deconvolutional layers for semantic segmentation in 2016. Several datasets including lesion segmentation and natural visual object segmentation are tested. Experiments show that the deconvolutional layers reduce the false positives, because these layers help to eliminate less discriminative features. 

FickleNet, a weakly or semi-supervised model for general image semantic segmentation, is proposed by \cite{lee2019ficklenet}. To achieve classification task to get the precise boundary with coarse annotation, it adopts to select hidden units randomly for calculating activation scores. In this way, the location information in the feature map are extracted and can be generated from a single image. In the end, location maps with region and objects information are used for training. \cite{peng2019discretely} came up a alternating direction method of multipliers (ADMM) algorithms to achieve training discretely-constrained neural networks in 2019. Providing weakly annotations including size constraints and boundary length, the segmentation for medical MRI images is achieved. Perone proposed a self-ensembling method for unsupervised domain adaptation\cite{perone2019unsupervised}.

\section{Reinforcement Learning for Segmentation}\label{section:RFL}
\begin{figure}[h]
    \centering
    \includegraphics[width=5.5cm]{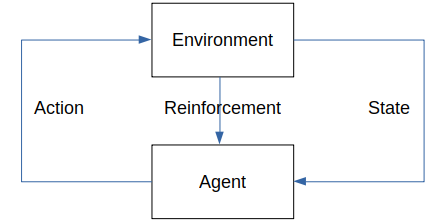}
    \caption{General model for a reinforcement learning agent}
    \label{fig:RFLFIGURE}
\end{figure}

Deep Reinforcement Learning (DRL) is a class of methods that uses neural networks as value function estimators. Its main advantage is using deep neural networks to extract state features automatically, avoiding manually defining state feature bands. This kind of inaccuracy makes the agent learn in a more primitive state. The general model for deep reinforcement learning agent is illustrated in Figure \ref{fig:RFLFIGURE}.

Sahba et al. introduced a Q-learning based method for ultrasound image segmentation in 2006 and 2008\cite{sahba2006reinforcement}\cite{sahba2008application}. A novel approach for setting local thresholding and structuring element values is introduced, and then the quality of segmented image can be used to fill a Q-metrix. Chitsaz et al. proposed a multi-object medical image segmentation framework in 2009\cite{chitsaz2009medical}. Each reinforcement agent is trained to find a optimal value for each object. Each state is associated defined actions, and punish/reward functions are calculated. Wang et al. came up a context-specific medical image segmentation framework with online reinforcement learning in 2013\cite{wang2013general}. To make the model be adaptive to prior knowledge/user's interaction, a general reinforcement learning which is suitable for large state-action space is proposed. Finally, this generic framework developed by user interaction and image content is tested to four different segmentation tasks.

\section{Evaluation Methods, Datasets and Results}\label{section:evaluation}
To collect the solid and objective performance results of each deep-learning-based approaches, comprehensive evaluation methods and public datasets are essential in the experiments. The section \ref{section:evaluation} reviews main evaluation methods of image segmentation, common datasets and summarize the results.
\subsection{Evaluation Method}
After the Convolutional Neural Network (CNN) was applied for visual data analysis, the performance results of proposed methods have to be evaluated and compared comprehensively. Commonly used performance evaluation indicators include average recall, pixel accuracy and intersection over union. To simply the expression of different evaluation methods, several parameters and notations are defined in detail. Without loss of generality, $k+1$ classes of target Organ $O$ areas (e.g. $O_{0}$ refers to a lung, $O_{1}$ refers to a kidney, $O_{k}$ refers to background and etc.), and the total of Pixels $P$ ($P_{ij}$ refers to the number of pixels organ $O_{i}$ are predicted to belong to the class of organ $O_{j}$) are considered. In other words, $P_{ii}$ refers to the true positive number of pixels. $P_{ij}$ refers to the false positive number of pixels. $P_{ji}$ refers to the false negative number of pixels. $P_{jj}$ refers to the true positive number of pixels.

\subsubsection{Pixel Accuracy}
Pixel accuracy is a ratio between the number of correctly classified pixels and the total number of pixels.
\begin{tiny}
\begin{equation}
PA=\frac{\sum_{i=0}^{k}P_{ii}}{\sum_{i=0}^{k}\sum_{j=0}^{k}P_{ij}}\label{eq}
\end{equation}
\end{tiny}
\subsubsection{Mean Pixel Accuracy}
Mean pixel accuracy is a average ratio for all classes of organs between the number of correctly classified pixels and the total number of pixels.
\begin{tiny}
\begin{equation}
MPA=\frac{1}{k+1}\sum_{i=0}^{k}\frac{P_{ii}}{\sum_{j=0}^{k}P_{ij}}\label{eq}
\end{equation}
\end{tiny}
\subsubsection{Pixel Precision}
Pixel precision is a ratio between the number of correctly classified positive pixels and the total number of positive predicted pixels.
\begin{tiny}
\begin{equation}
PP=\frac{\sum_{i=0}^{k}P_{ii}}{\sum_{i=0}^{k}(P_{ii}+P_{ij})}\label{eq}
\end{equation}
\end{tiny}
\subsubsection{Mean Pixel Precision}
Mean pixel precision is a average ratio for all classes of organs between the number of correctly classified positive pixels and the total number of positive predicted pixels.
\begin{tiny}
\begin{equation}
MPP=\frac{1}{k+1}\sum_{i=0}^{k}\frac{P_{ii}}{P_{ii}+P_{ij}}\label{eq}
\end{equation}
\end{tiny}
\subsubsection{Pixel Recall}
Pixel recall is the proportion of boundary pixels in the ground truth that were correctly classified by the segmentation.
\begin{tiny}
\begin{equation}
PR=\frac{\sum_{i=0}^{k}P_{ii}}{\sum_{i=0}^{k}P_{ii}+\sum_{j=0}^{k}P_{jj}}\label{eq}
\end{equation}
\end{tiny}
\subsubsection{Dice Coefficient}
Dice coefficient, also known as Sørensen–Dice index, is a statistic used to gauge the similarity of two boundary.
\begin{tiny}
\begin{equation}
DSC=\frac{2*\sum_{i=0}^{k}P_{ii}}{2*\sum_{i=0}^{k}P_{ii}+\sum_{i=0}^{k}\sum_{j=0}^{k}(P_{ij}+P_{ji})}\label{eq}
\end{equation}
\end{tiny}
\subsubsection{Conformity Coefficient}
Conformity coefficient is developed based on Dice coefficient which can also be used to discriminate between surface, volume and boundary.
\begin{tiny}
\begin{equation}
CF=\frac{3*DSC-2}{DSC}\label{eq}
\end{equation}
\end{tiny}
\subsubsection{Intersection Over Union}
Intersection over union, also known as Jaccard index, is the percent overlap between the target mask and the prediction output.
\begin{tiny}
\begin{equation}
IOU=\frac{\sum_{i=0}^{k}P_{ii}}{\sum_{i=0}^{k}\sum_{j=0}^{k}P_{ij}-\sum_{j=0}^{k}P_{jj}}\label{eq}
\end{equation}
\end{tiny}
\subsubsection{Mean Intersection Over Union}
Mean Intersection over union is the average percent overlap for each class between the target mask and the prediction output.
\begin{tiny}
\begin{equation}
MIOU=\frac{1}{k+1}\sum_{i=0}^{k}\frac{P_{ii}}{\sum_{j=0}^{k}P_{ij}+\sum_{j=0}^{k}P_{ji}-P_{ii}}\label{eq}
\end{equation}
\end{tiny}
\subsubsection{Frequency Weighted Intersection Over Union}
Frequency weighted Intersection over union is the average percent overlap for each class importance depending on appearance frequency between the target mask and the prediction output.
\begin{tiny}
\begin{equation}
FWIOU=\frac{1}{\sum_{i=0}^{k}\sum_{j=0}^{k}P_{ij}}\sum_{i=0}^{k}\frac{\sum_{j=0}^{k}P_{ii}P_{ij}}{\sum_{j=0}^{k}P_{ij}+\sum_{j=0}^{k}P_{ji}-P_{ii}}\label{eq}
\end{equation}
\end{tiny}

\subsection{Dataset}
To evaluate each segmentation method fairly and objectively, an authoritative dataset plays an essential role in evaluation. Table \ref{tab:Datasets} summarizes the common public datasets for ultrasound image segmentation.
\subsection{Experimental Comparison of Segmentation Accuracy}
To illustrate the contributions and evaluation results for different methods, Table \ref{tab:Results} summarizes each method according to the category, contributions, datasets and results.

\begin{table*}[h]
\tiny
	\centering  
	\caption{Common Datasets for Medical Ultrasound Image}  
	\label{table1} 
	\begin{tabular}{c c c c c}  
		\hline 
		\textbf{Dataset}  & \textbf{Resolution} & \textbf{Size of set} & \textbf{Description} & \textbf{Device}\\ 
		\hline
		
		Ultrasound Nerve Segmentation Challenge & 580$\times$420 & 5636 & A collection of nerves called the Brachial Plexus (BP) & N/A \\
		 %&&&&\\
% 		 \cite{datasetUNSC}
		
		\multirow{2}{*}{Breast Ultrasound Teaching File} & \multirow{2}{*}{Average 377$\times$396} &  \multirow{2}{*}{6600} & Age, Menopausal, Hormonal Replacement, & \multirow{2}{*}{B\&K Medical System} \\
		&&&Family History Physical Exam are provided&\\
% \cite{prapavesis2003breast}

		\multirow{2}{*}{Breast Ultrasound Lesions Dataset} & \multirow{2}{*}{Average 760$\times$570} & \multirow{2}{*}{469} & 306 (60 malignant and 246 benign) & \multirow{2}{*}{Siemens ACUSON Sequoia C512}  \\
		&&& 163 (53 malignant and 110 benign)&\\
% \cite{yap2017automated}
		
		\multirow{2}{*}{Malignant and Benign Breast Lesions} & \multirow{2}{*}{N/A} & \multirow{2}{*}{100} & \multirow{2}{*}{78 women aged from 24 to
75} & Ultrasonix SonixTouch  \\
		&&&&Research ultrasound scanner\\
% 		\cite{piotrzkowska2017open}

		\multirow{2}{*}{Malignant Solid Mass} & \multirow{2}{*}{200$\times$200-300$\times$400} & \multirow{2}{*}{241 cases} &  Labeled by three experts by the majority &  \multirow{2}{*}{A Philips iU22 ultrasound machine}  \\
		&&& voting rule (two out of three)&\\
% 		\cite{datasetOMI}
		
		\multirow{2}{*}{Cervical Dystonia} & \multirow{2}{*}{491$\times$525} & \multirow{2}{*}{3272} & \multirow{2}{*}{61 adults: 35 cervical dystonia and 26 normal} &  Probe (7.5MHz, SonixTouch,  \\
		&&&& Ultrasonix, USA)\\
% 		\cite{loram2020objective}
		
		Biometric Measurements from Fetal Ultrasound Images & N/A & 284 & 14 fetus, 90 femur, 90 head, 90 abdomen & N/A \\
		%&&&&\\
% 		\cite{datasetOXFUI}
		
		\multirow{2}{*}{Fetal US Image Segmentation} & \multirow{2}{*}{756$\times$546} & \multirow{2}{*}{9 Training set, 1 Test set} & Fetal head, abdomen, & \multirow{2}{*}{Philips HD9} \\
		&&& and femur sub-challenges&\\
% 		\cite{rueda2013evaluation}
		
		Vessel Segmentation & N/A & TBU & Released per month with over 7000 cases & GE Voluson US imaging system \\
		%&&&&\\
% 		\cite{Vessel}

		\multirow{2}{*}{IVUS Challenge, MICCAI} & \multirow{2}{*}{up to \SI{113}{\micro\metre}} & \multirow{2}{*}{512 $\times$ 5 frames} & Outer wall of media and adventitia segmentation, & \multirow{2}{*}{40 MHz IVUS scanner} \\
		&&&inner wall of lumen segmentation&\\
% 		\cite{balocco2014standardized}

		\multirow{2}{*}{SYSU-FLL-CEUS} & \multirow{2}{*}{768$\times$576} & \multirow{2}{*}{10 videos} &Three types: 186 HCC, & Aplio SSA-770A\\
		&&&109 HEM and 58 FNH instances & (Toshiba Medical System)\\
% 		\cite{liang2015recognizing}
		
		\hline
	\end{tabular}
	\label{tab:Datasets}
\end{table*}

\begin{table*}
\tiny
    \centering
    \caption{Comparison of Medical Ultrasound Image Segmentation Results}
    \begin{tabular}{c c c c c}
         \hline 
         \textbf{Category} &  \textbf{Model}  & \textbf{Contributions} & \textbf{Dataset} & \textbf{Evaluation}\\
         \hline
         \multirow{7}{*}{\textbf{FCN}} & \multirow{2}{*}{CFS-FCN\cite{zhang2016coarse}}  & Multi-stage incremental learning & \multirow{2}{*}{80 ultrasound images} & \multirow{2}{*}{MIOU: 5\%} \\
                                                              & & Several small-size FCN & & \\
                                                              & \multirow{2}{*}{CasFCN\cite{wu2017cascaded}} & Cascaded framework & \multirow{2}{*}{900 fetal head and 688 abdomen images} & \multirow{2}{*}{DSC: 0.9843} \\
                                                              & & Auto-Context scheme  & & \\
                                                              & DSN-OB\cite{mishra2018ultrasound} & Auxiliary losses for boundary detection & MICCAI 2011 IVUS challenge & DSC: 0.91 \\
                                                              & FCN-TN\cite{li2018fully} & Dropout to prevent overfitting  & 300 Thyroid Nodules US images & IOU: 91\% \\
                                                              & DF-FCN\cite{yang2019improving} & A catheter detection method to reorganized cross-sections  & 25 3D US dataset & DSC: 57.7\% \\
         \hline
         \multirow{8}{*}{\textbf{En-Decoder}} & DeconvNet+ \cite{cunningham2019ultrasound} & Completely symmetric structure with Maxout & 1100 transversal neck US images of 28 adults & JI: 0.532, HD: 5.7mm\\
                        & Dynamic CNN\cite{zhou2019deep} & Dynamically fine tuned &  144 carotid 3D ultrasound & DSC: 0.928--0.965\\
                                                          & GRA U-Net\cite{zhuang2019nipple} & Modified U-Net to Grouped-Resaunet segment &  25 patients with 131 slices & MIOU: 0.847\\
                                                           & \multirow{2}{*}{\cite{kim2018fully}} & Multi-label loss function& \multirow{2}{*}{IVUS dataset} & \multirow{2}{*}{DSC: 0.84}\\
                                        & & Weighted pixel-wise cross-entropy& & \\                 &NAS-Unet\cite{weng2019unet} & Network architecture search & Kaggle Ultrasound nerve segmentation & MIOU: 0.992\\
                                                          & \multirow{2}{*}{DS-CR-V-Net\cite{lei2019ultrasound}} & 3D convolution,residual blocks & \multirow{2}{*}{44 patients' TRUS data} & \multirow{2}{*}{DSC: 0.919}\\
                                        & &Multi-directional deep supervision& & \\
         \hline
         \multirow{3}{*}{\textbf{RNN}} & T-RNN \& J-CNN\cite{chen2015automatic} &  Knowledge transferred recurrent neural networks &  300 videos with 37376 US images & AUC: 0.95 \\
                                                          & 3D-FCN\cite{yang2017towards} \& RNN & Deep supervision for volumetric segmenta
                                                          tion &  104 anonymized prenatal ultrasound volumes & DSC: 0.882\\
                                                          & Multi-view BCRNN\cite{yang2017fine} & Utilize shape prior to infer the boundary &  17 trans-rectal ultrasound volumes & DSC: 0.9239\\
        \hline
         \multirow{4}{*}{\textbf{GAN}} & PAN\cite{khosravan2019pan} & Address computational burden & TCIA dataset & DSC:86.8\\
                                                          & SegAN\cite{xue2018segan} & Multi-scale loss function & MICCAI BRATS & DSC: 0.85, Precision: 0.92\\
                                                          & \multirow{2}{*}{\cite{yang2018generalizing}} & Convert the target corpus on the-fly &  \multirow{2}{*}{2699 US Images} & \multirow{2}{*}{DSC: 93.379}\\
                                                          & & A self-play training strategy & & \\
        \hline
         \multirow{5}{*}{\textbf{WSL}} &\multirow{2}{*}{\cite{kim2016deconvolutional}} & \multirow{2}{*}{Unpooling-deconvolution networks} &  MC (80 normal, 58 abnormal) & \multirow{2}{*}{IOU: 21.61, 24.61} \\
                                                          & & & Shenzhen (326 normal, 336 abnormal) & \\
                                                          & FickleNet\cite{lee2019ficklenet} & Stochastic selection of hidden layers  &  PASCAL VOC 2012 & MIOU: 61.2\\
                                                          & \cite{peng2019discretely} & Training with discrete constraints& Automated Cardiac Diagnosis Challenge& Mean DSC: 0.901\\    
                                                & \cite{perone2019unsupervised} & Unsupervised domain adaptation & Spinal Cord Gray Matter Challenge & DSC: 84.72\\          
        \hline
         \multirow{3}{*}{\textbf{DRL}} & \cite{sahba2008application} & Utilize local to extract the prostate &  60 images with specific prostate shape & Area Overlap: 0.9096\\
                                                          & \cite{chitsaz2009medical} &Different ROIs segmentation & A series
of 512$\times$512 medical images & Accuracy: 0.93\\
                                                          &\cite{wang2013general}& Context-specific segmentation & LV, RV Segmentation Dataset & Jaccard Index: 0.922\\                          
         \hline  
         
    \end{tabular}
    \label{tab:Results}
\end{table*}

\section{Discussion and Potential Directions}\label{section:DISCUSSION}
Deep learning was widely applied to medical image segmentation in the past decade. This review mainly focuses on deep learning approaches for medical ultrasound image segmentation. Typical methods are categorized into six groups, introduced and summarized. Based on the existing researches and studies, this review summarizes key issues, challenges and some potential research directions.
\\$\bullet$ In the last five years, deep learning has demonstrated state-of-the-art performance in ultrasound image segmentation tasks. Nearly 80\% deep learning models, however, are developed based on 2D fully convolutional neural networks or Encoder-decoder networks. These approaches request high cost in labeling data and can not study on the sequence of ultrasound images or the position of ultrasound scanners, which results in the loss of context information. Furthermore, clinicians or other health-workers normally labels ultrasound images with bounding boxes, lines or object types on the most distinctive image rather than pixel-level boundary annotation, so the shortcomings of accurately labeling datasets remain a challenge. Therefore, there is still a lot of study work to do on applying 3D fully convolutional networks, recurrent neural networks, generative adversarial networks, weakly supervised learning to make full use of current public datasets for ultrasound image segmentation.
\\$\bullet$ Segmentation performance is usually evaluated by accuracy rather than speed or memory cost. With the development of medical hardware, the ultrasound equipment is developed toward miniaturization, high efficiency and ease of daily use with a portable computer. Therefore, the machine learning model should be as lightweight as possible to save more time while ensuring accuracy and stability. It can be promising once the model is deployed on a entry-level CPU.
\\$\bullet$ The heterogeneous appearance of the organ is one of the biggest challenges in ultrasound image segmentation. The pattern of ultrasound images can be different varies between the organ location, depth, neighboring tissues, hardness even the operators. Several non-machine learning approaches such as deformable models, watershed, region grow and graph-based methods are essential for general ultrasound image segmentation tasks. The study of deep learning can be potentially inspired by traditional methods, and the evaluation should be proved in general segmentation tasks. 
\\$\bullet$ In clinical medical imaging, there are a number of multiple modalities methods such as MRI, X-ray, PET and CT. The medical imaging methods for diagnostic are adopted depending on different cases. Ultrasound image is regarded as the first-line method because of its low cost and non-radiation. MRI or CT is time consuming, costly but rich in texture. To solve the processing work for multi resources, cross-modal transfer learning can be a potential research direction.
\\$\bullet$ From the perspective of clinical diagnosis, on the one hand, the intrinsic speckle noises may affect the imaging effects to a certain extent. On the other hand, some small organ structures may be obscured by large human organs, so restoring obstructed lesions to reduce missed diagnosis is a challenge. Furthermore, early prediction of diseases is the potential research direction in the future. The survival rate of patients will be greatly improved, if the diseases can be detected earlier, but all of these deep-learning-based models are trained on the datasets labeled by doctors, the features failed to be labeled by doctors can not be learned by the training models. To this end, training models with the labeled data and then expanding knowledge to make advance diagnosis prediction is of great significance in the future. 

%  \section*{Acknowledgment}

%  The authors would like to thank Nanqing Dong for his valuable suggestions.

% \small
\bibliographystyle{named}
\bibliography{ijcai20}

\end{document}